\documentclass[openacc]{rstb}

\usepackage{graphicx}
\usepackage{xcolor}
\usepackage{amsmath}
\usepackage{braket}

\newcommand{\up}{\uparrow}
\newcommand{\down}{\downarrow}

\titlehead{Research}

\begin{document}
\title{Modeling the quantum-like dynamics of human reliability ratings in Human-AI interactions by interaction dependent Hamiltonians}

\author{Johan van der Meer, Pamela Hoyte, Luisa Roeder, Peter Bruza}

\address{School of Information Systems, Queensland University of Technology, Brisbane, Australia\\
}
\subject{Quantum Cognition, Decision Making, Trust}

\corres{Peter Bruza\\
\email{p.bruza@qut.edu.au}}

\begin{abstract}
As our information environments become ever more powered by artificial intelligence (AI), the phenomenon of trust in a human's interactions with this intelligence is becoming increasingly pertinent. 
For example, in the not too distant future, there will be teams of humans and intelligent robots involved in dealing with the repercussions of high-risk disaster situations such as hurricanes, earthquakes, or nuclear accidents. Even in such conditions of high uncertainty, humans and intelligent machines will need to engage in shared decision making, and trust is fundamental to the effectiveness of these interactions.

A key challenge in modeling the dynamics of this trust is to provide a means to incorporate sensitivity to fluctuations in human trust judgments. In this article, we explore the ability of Quantum Random Walk models to model the dynamics of trust in human-AI interactions, and to integrate a sensitivity to fluctuations in participant trust judgments based on the nature of the interaction with the AI.

We found that using empirical parameters to inform the use of different Hamiltonians can provide a promising means to model the evolution of trust in Human-AI interactions.
\end{abstract}
\keywords{trust, AI, quantum modeling}
\rsbreak

\section{Introduction}

A well-cited conception of trust is based on three features: Ability, Benevolence and Integrity \cite{mayer:1995}.
Ability is the trust in the skills, competencies, and characteristics of the trustee, e.g., the intelligent robot. 
Benevolence is the belief of goodness and goodwill perceived in the trustee, 
and Integrity is the adherence to a set of principles or 
values that the trustor finds acceptable. 
Placing these characteristics in the context of human-AI interactions,
trust involves the human’s perception of the AI’s competence (functional capability) and reliability (performance integrity, consistency with expectations) of the AI to perform its duties, as well as positive affect alignment (belief that the AI will function in the best interest of the human and their goals).

Lurking behind this conception of trust is the assumption that the degree of trust is causally dependent on the degree of presence of these three features. 
Consistent with this line of thinking is that trust involves cognitive states that have a definite value at all times, because causality relies on states of reality that are determinate.
By way of illustration, imagine a human agent interacting with an AI system where the reliability of the AI is judged on a scale of ``low", ``average", ``high". Based on the behavior of AI, these judgments may change over time.
In order to model the dynamics of these reliability judgments, an orthodox cognitive model would assume that, at all points in time, a human agent's cognitive state would hold one of these three values, and that this value would simply be reported when the human agent is probed about the AI's reliability. 
In other words, this model is based on the assumption that the cognitive decision state has a definite value at all times, even during periods when the human agent is not being probed about the AI's reliability. 

Consequently, in this orthodox view, the dynamics of the reliability judgments can be modeled by a Markov process whereby at each time point there is a probability associated for transition of a given state to another state. For example, assuming that the current cognitive state reflects a judgment that the AI has ``low reliability", there are probabilities of remaining with that judgment and of transitioning to judgments corresponding to ``average" and ``high".
In this way, a Markov model can predict the probability that human agents hold a given reliability judgment at an arbitrary time $t$. In addition, the dynamics of the human-AI interactions can be viewed as a random walk (RW) through a network of nodes, where each node corresponds to a definite cognitive state regarding a specific judgment of the AI's reliability at a given time.

An alternative position we adopt in this paper is to view trust as an \emph{indeterminate} cognitive state as proposed by the field of quantum cognition \cite{busemeyer:bruza:2012, pothos2022quantum, busemeyer:bruza:2025}.
Under this view, there is no definite value of the cognitive state before the human agent is probed to report his or her judgment of the AI's reliability. Only when probed does one of the values actualize at the point of measurement.
This view parallels the situation with quantum phenomena, the properties of which are assumed to indeterminate before measurement. What this means is that the values of a property correspond to ``potentialities" \cite{Jaeger:2017}, which can be considered as propensities that are only made manifest by the act of measurement.

\begin{figure}[h!]
\centering
\includegraphics[width=0.5\columnwidth]{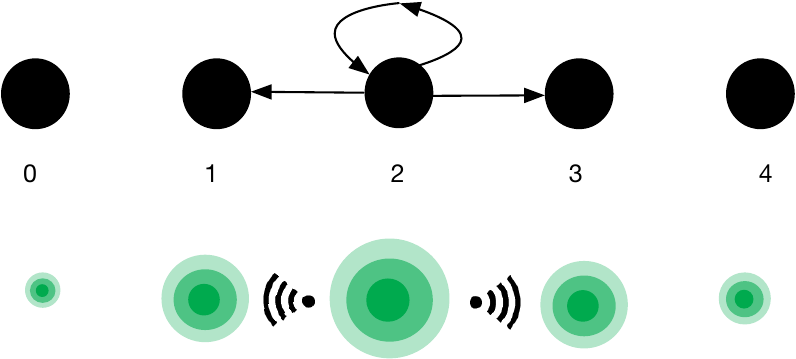}
\caption{Markov and quantum models. The~top depicts a Markov model. The~black dots represent definite cognitive states associated with a reliability rating. The~human agent currently inhabits the cognitive state associated with a reliability rating of 2. In~a RW walk model, there can be a transition to adjacent states or back to the current state. Probabilities are associated with these transitions. The~quantum model (bottom) illustrates that the human agent simultaneously inhabits all cognitive basis states which are like wave peaks. The~concentric circles represent the amplitude of the wave peaks at each basis state, which represent probabilities. Wave energy transfers to neighbouring wave~peaks.}
\label{fig:quantum-markov}
\end{figure}

The dynamics of the reliability judgments under the assumption of indeterminate cognitive states can be modeled by a Quantum Random Walk model (QRW) model. In contrast to the RW described above, where the dynamics correspond to a path through the network, quantum dynamics can be envisaged as all paths being traversed simultaneously, like a wave with multiple crests dynamically emanating through the network, until the point of measurement. Associated with each path is an amplitude by which, just like waves, these amplitudes can interfere with each other and this interference attenuates the underlying probabilities. This dynamicism is a significant and defining feature that differentiates predictions from quantum models in comparison to models assuming definite states, such as the Markov model described previously \cite{busemeyer:bruza:2012}. 
This difference between the Quantum and Markov models is illustrated in Figure~\ref{fig:quantum-markov}.

\vspace{-4pt}

\begin{figure}[h!]
	\centering	\includegraphics[width=0.7\columnwidth]{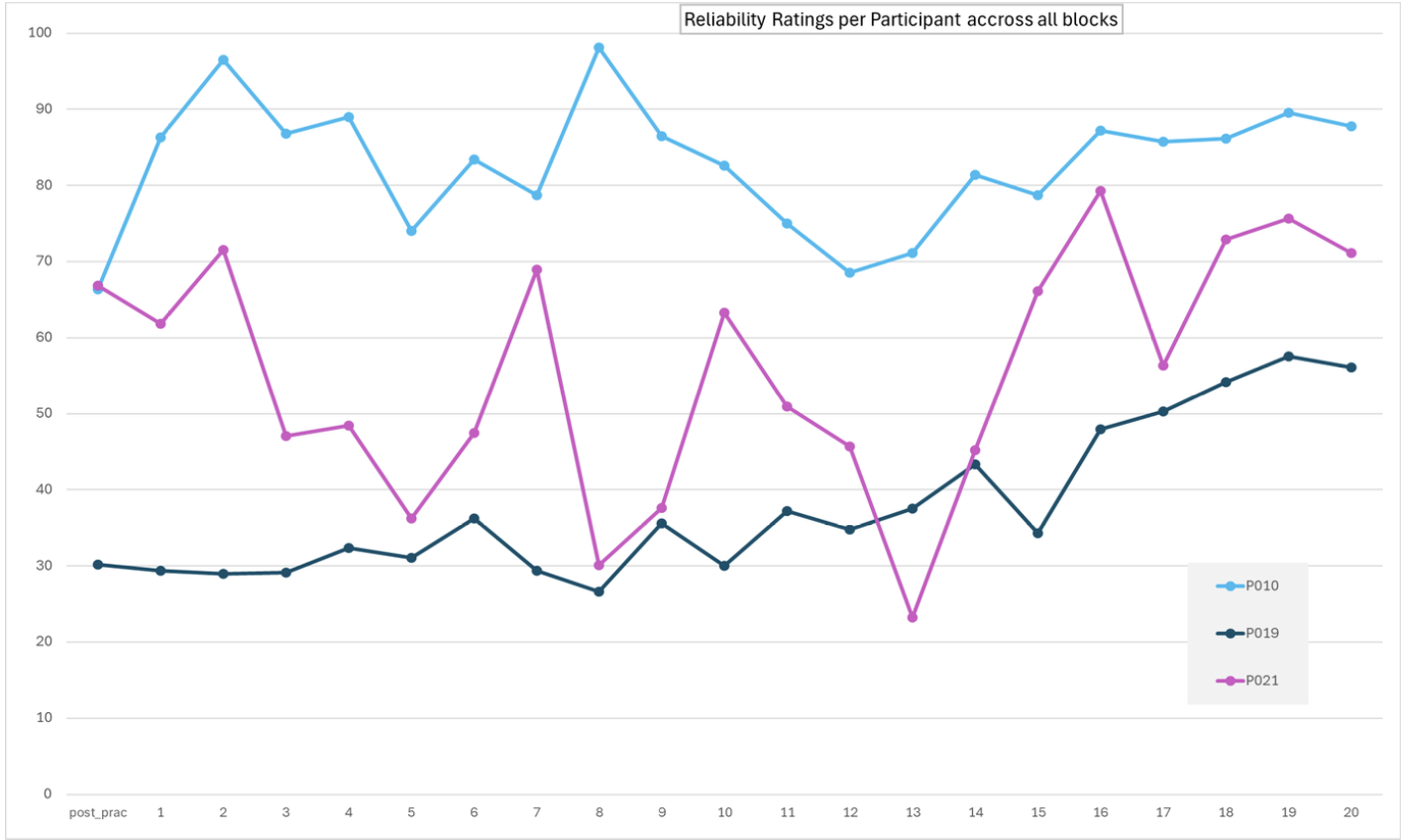}
        \caption{Variation of reliability ratings of AI for three participants}
\label{fig:reliability-ratings}
\end{figure}

Although QRW models have shown promise in modelling the dynamics of human reliability ratings of AI \cite{roeder2023quantum} and trust judgments of AI \cite{canan2021addressing}, 
a main challenge is catering for the fluctuations of ratings both across and within participant/ interactions with the AI.
For example, Figure \ref{fig:reliability-ratings} exhibits such fluctuations for three participants in a study that we present in more detail below.

In order to help ground the model proposed below, an experiment is described, which illustrates the conceptual framing of the model. 
In the experiment \cite{roeder2023quantum}, human participants were required to interact with a Wizard-of-Oz (WoZ) AI system for an image classification task, and periodically rate the reliability of the AI system.
Specifically, participants were shown images of human faces one by one and given up to 2 seconds to judge whether it was real or fake (that is, GANs generated).
After the participants recorded their judgment, they were shown the AI classification of the image as real or fake. If they did not record a judgment within 2 seconds, 'no response' was recorded and the image classification task moved on to the next trial.
The WoZ element is that the judgments of the AI system were manipulated to often agree (match) with the participant's response (randomized 75\% match-25\% mismatch).
This frequent agreement is intended to promote the expectation that the AI is reliable.
When the~AI system disagrees (mismatches) with the human response, the expected reliability that has been promoted is perturbed. Participants rated the reliability of the AI system on a hundred point scale (0-100) at regular intervals (after blocks of 28 trials). In total, after an initial practice block, there were 20 blocks of 28 trials, with a short breaks after block 7 and 14. Data were collected from 34 participants, including the reliability ratings for the 3 participants depicted in Figure \ref{fig:reliability-ratings}. 
All experimental protocols were approved by the Human Research Ethics Committee of Queensland University of Technology (LR 2022-5210-10103, 15 June 2022) and by the U.S. Air Force Human Research Protection Office (FWR20220251X, 1 November 2022) in accordance with the Declaration of Helsinki.

In order to better understand the challenge posed by the fluctuations, such as depicted in Figure \ref{fig:reliability-ratings}, we first expand on the nature of QRW dynamics.

\subsection{Dynamics of the Quantum Random Walk model}

The QRW dynamics is prescribed by a $n \times n$ Hamiltonian matrix $H$, where $n$ is the dimensionality of the Hilbert space. For~example, when $n=5$,
\begin{align*}
H &= \begin{bmatrix}
     \mu_0 & \sigma^2 & 0 & 0 & 0 \\
     \sigma^2 & \mu_1 & \sigma^2 & 0 & 0\\
     0 & \sigma^2 & \mu_2 & \sigma^2 & 0 \\
     0 & 0 & \sigma^2 & \mu_3 & \sigma^2 \\
     0 & 0 & 0 & \sigma^2 & \mu_4
     \end{bmatrix}
\end{align*}
The value in cell $h_{ij}$ specifies the diffusion of the wave amplitude from the column state $j$ to the row state $i$. 
Therefore, the~parameter $\mu_x$ specifies the diffusion rate back into a basis state $\ket{x}$. In~other words, it corresponds to energy for the human agent to keep expectations of reliability at a certain level in relation to a reliability rating $x$.
The diagonal values correspond to the nature of the potential function, which drives the dynamics.

When the diagonal values are increasing, 
e.g., $\mu_x = \beta x$ specifies that a constant positive force is being exerted, the energy of the wave drives towards higher reliability ratings $x$.
This occurs when the AI system is conforming to expectations.
In contrast, the~parameter $\sigma^2$ specifies the diffusion of wave amplitude away from a given reliability state to neighbouring reliability states.
The non-zero off-diagonal values correspond to wave energy dissipating to neighbouring reliability levels. 
The dynamics of the quantum model are defined via a unitary matrix using the Schr\"odinger equation:
\begin{align}
\frac{d}{dt}U(t) &= -iHU(t)\\
U(t) &= e^{-itH} 
\end{align}
The evolution of the cognitive state $\psi$ at time $t$ is given via:
\begin{align}
\frac{d}{dt}\psi(t) &= -i H \psi(t), \\
\psi(t) &= e^{-itH}\psi(0) \label{formula:unitary-evololution}
\end{align}
Note, however, that as shown in Figure \ref{fig:reliability-ratings} empirical reliability ratings increase and decrease, sometimes dramatically. Retaining a positive $\beta$ will produce dynamics that simply cannot model such fluctuations.

The goal of this article is to provide a quantum-like dynamics that is more sensitive to fluctuations in reliability ratings.
The key idea proposed to enable this is that when the AI responses align with human judgment, the expectation grows that the AI is reliable. In such cases, $\beta$ should be positive.
Conversely, when the AI responses do not align with human judgment, the expectation grows that the AI is unreliable.
Therefore, $\beta$ should be negative.
In this way, the Hamiltonian used is dependent on the nature of the underlying interaction, i.e., alignment or misalignment. 
A key question to be addressed is when to appropriately update $\beta$ in the Hamiltonian.

\section{A Quantum Random Walk Model based on interaction dependent Hamiltonians}


In our previous work \cite{roeder2023quantum}, the QRW model prescribed the dynamics using a Hamiltonian with three parameters and did not incorporate events that occurred during the experiment.  
The potential disadvantage of this approach is that ignoring events and circumstances that occur during  human-AI interaction which might affect judgments of reliability, served to dampen the model's dynamicism and sensitivity to fluctuations. The QRW model proposed here expands on this approach by incorporating the events, times, and choices that occurred during the experiment.
Specifically, these considerations influence 
1) how the cognitive state vector is defined, 
2) when the state vector is collapsed, 
3) the time-steps per trial, and 
4) the result of the interaction with the AI (match, mismatch or no response).

Firstly, we modeled reliability levels with a cognitive state vector $\psi$ consisting of 10 levels. The first level (0.5) represents 0-10\% trust, the second (1.5) represents 10-20\%, and so on up to the 10th level, with 9.5 representing 90-100\% trust. The state vector at time $t$ is therefore represented as:
\[
\psi(t) =
\begin{bmatrix}
\psi_1(t) \\
\psi_2(t) \\
\psi_3(t) \\
\vdots \\
\psi_{10}(t)
\end{bmatrix}
\label{formula:init_state_vec}
\]
The initial state is determined by the participant's reliability rating after the practice block. The amplitudes of the initial state are set according to a Gaussian curve of mean \( \text{reliability rating} / 10 \) and standard deviation 0.75, sampled at 0.5, 1.5, ..., 9.5. The amplitudes are rescaled to ensure that their squares sum to unity.

Secondly, we postulate the collapse of the cognitive state after each block at the point the participant rates the AI's reliability. During the trial blocks, trust is an indeterminate state evolving during the participant's image classification decisions and their perception of the AI agent's decisions. After 28 such evolutions (trials), the participant answers the question "How reliable do you find the AI rating?". We propose that this question causes the participant to consider their internal state of trust, collapsing the indeterminate state, and pausing its evolution until a new block (of 28 trials) starts again. In order to collapse the state, we set the amplitudes according to a Gaussian distribution that is centred on the reported reliability rated supplied by the participant, sampled at the intervals according to \eqref{formula:init_state_vec}, whereby the phases are kept intact.

Third, the time evolution per trial is modeled with the time difference between consecutive trials. Only for the first trial in each block, we use the median time difference (across all trials) as the time step. Since we do not know exactly the correct rate of time evolution as defined in Equation \eqref{formula:unitary-evololution}, we scale this evolution speed with a parameter $\gamma$, which has been referred to as an ``attenuation parameter" \cite{kvam2015interference}:
\begin{align}
\psi(t) &= e^{-itH\gamma}\psi(0) 
\end{align}
For our modeling, we explore values for $\gamma$ ranging from 0.01 to 1. Because we enter the actual time difference into the model and use three different Hamiltonians, we need this extra parameter, to act as an affective accelerator or brake, thereby controlling how far the quantum state vector is evolved with each time step.


Fourth, we use the property that the diagonal of the Hamiltonian induces a gradient to evolve the state with different Hamiltonians for different trials: for match trials, we use a Hamiltonian with a positive gradient, for mismatch trials, we use a Hamiltonian with a negative gradient. For 'no response' trials, we use a Hamiltonian with a constant value throughout the diagonal.

\[
\mathbf{H}_{+} =
\begin{bmatrix}
-4.5\alpha & \sigma & 0 & \cdots & 0 \\
\sigma & -3.5\alpha & \sigma & \cdots & 0 \\
0 & \sigma & -2.5\alpha & \cdots & 0 \\
\vdots & \vdots & \vdots & \ddots & \vdots \\
0 & 0 & 0 & \cdots & 4.5\alpha
\end{bmatrix}
\]
\[
\mathbf{H}_{-} =
\begin{bmatrix}
4.5\beta & \sigma & 0 & \cdots & 0 \\
\sigma & 3.5\beta & \sigma & \cdots & 0 \\
0 & \sigma & 2.5\beta & \cdots & 0 \\
\vdots & \vdots & \vdots & \ddots & \vdots \\
0 & 0 & 0 & \cdots & -4.5\beta
\end{bmatrix}
\]
\[
\mathbf{H}_{0} =
\begin{bmatrix}
1 & \sigma & 0 & \cdots & 0 \\
\sigma & 1 & \sigma & \cdots & 0 \\
0 & \sigma & 1 & \cdots & 0 \\
\vdots & \vdots & \vdots & \ddots & \vdots \\
0 & 0 & 0 & \cdots & 1
\end{bmatrix}
\]

In these three Hamiltonians, \( \alpha \), \( \beta \), and \( \sigma \) are modeling parameters. \( \alpha \) and \( \beta \) model slopes, and \( \sigma \) models the diffusion. For our modeling, we explore values for these three parameters ranging from 0.01 to 10.

\begin{figure}
    \centering
    \includegraphics[width=0.8\textwidth]{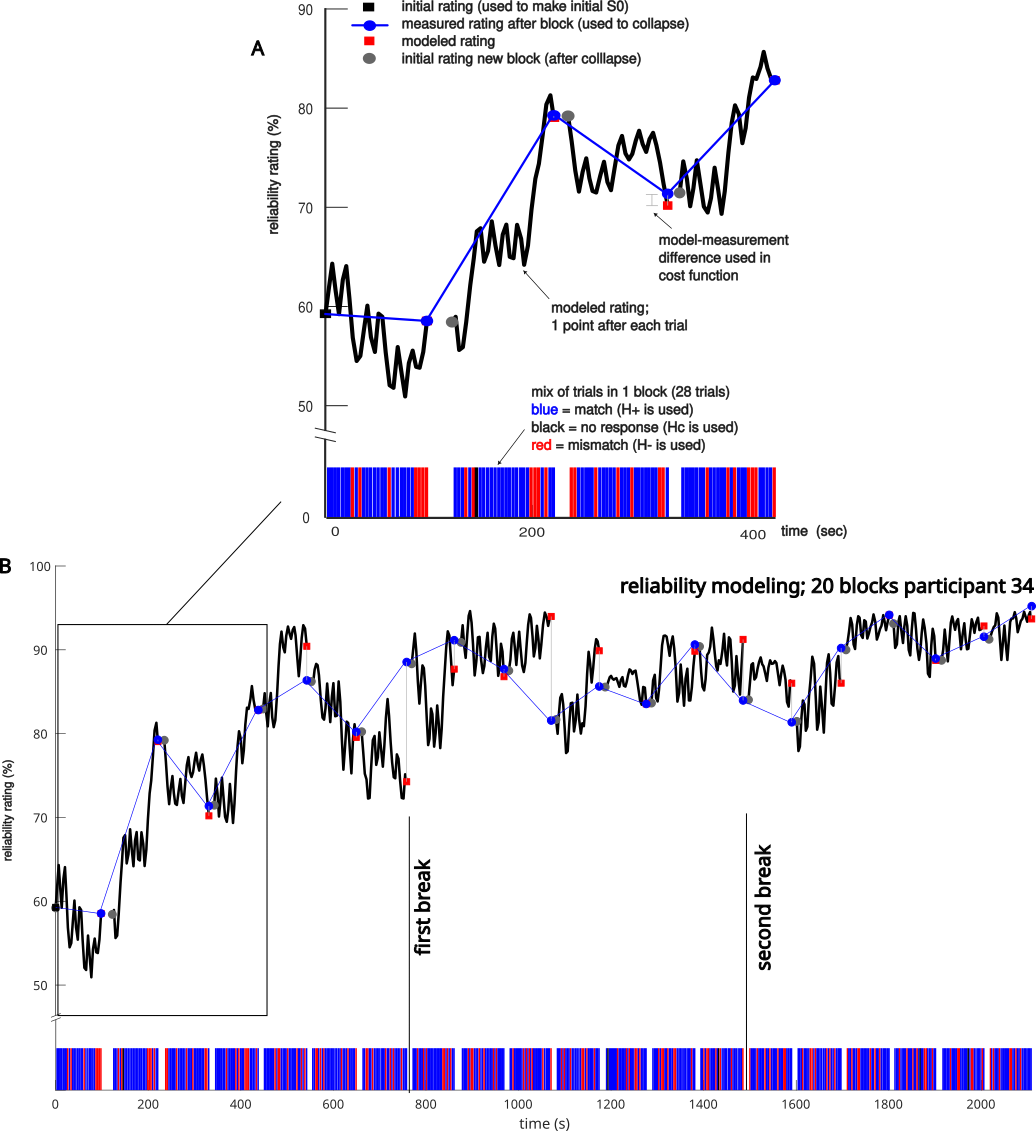}
    \caption{Reliability modeled with the Quantum Random Walk Model, adopted to our WoZ Trust experiment. Panel A: The  cognitive state vector is centred around the initial reliability rating reported by the participant; the mix of Hamiltonians (positive, negative and neutral) affect the evolution of the state throughout the first block. After finishing 28 trials, the participant rates the AI reliability, the state collapses and is 'frozen' until the start of the second block. The difference between the modeled and human-provided reliability rating (blue line) is used to find optimal model parameters for this participant. Panel B: the full experiment has 20 blocks of 28 trials each; these 20 blocks are divided into three parts with short breaks.}
    \label{fig:results_34_and_explanation}
\end{figure}

Using these four modeling assumptions, the data (types of trials, times, initial rating, and reliability ratings), we search the parameter space given the boundary conditions above. In order to find optimal parameters for \( \gamma, \alpha, \beta, \sigma \), we performed a global optimization using mean absolute error (MAE) as the cost function, comparing the modeled and observed reliability ratings. We have run this optimization for all 32 participants in our study and present a typical example of participant 34 in Figure \ref{fig:results_34_and_explanation}.

\begin{figure}
    \centering
    \includegraphics[width=0.8\textwidth]{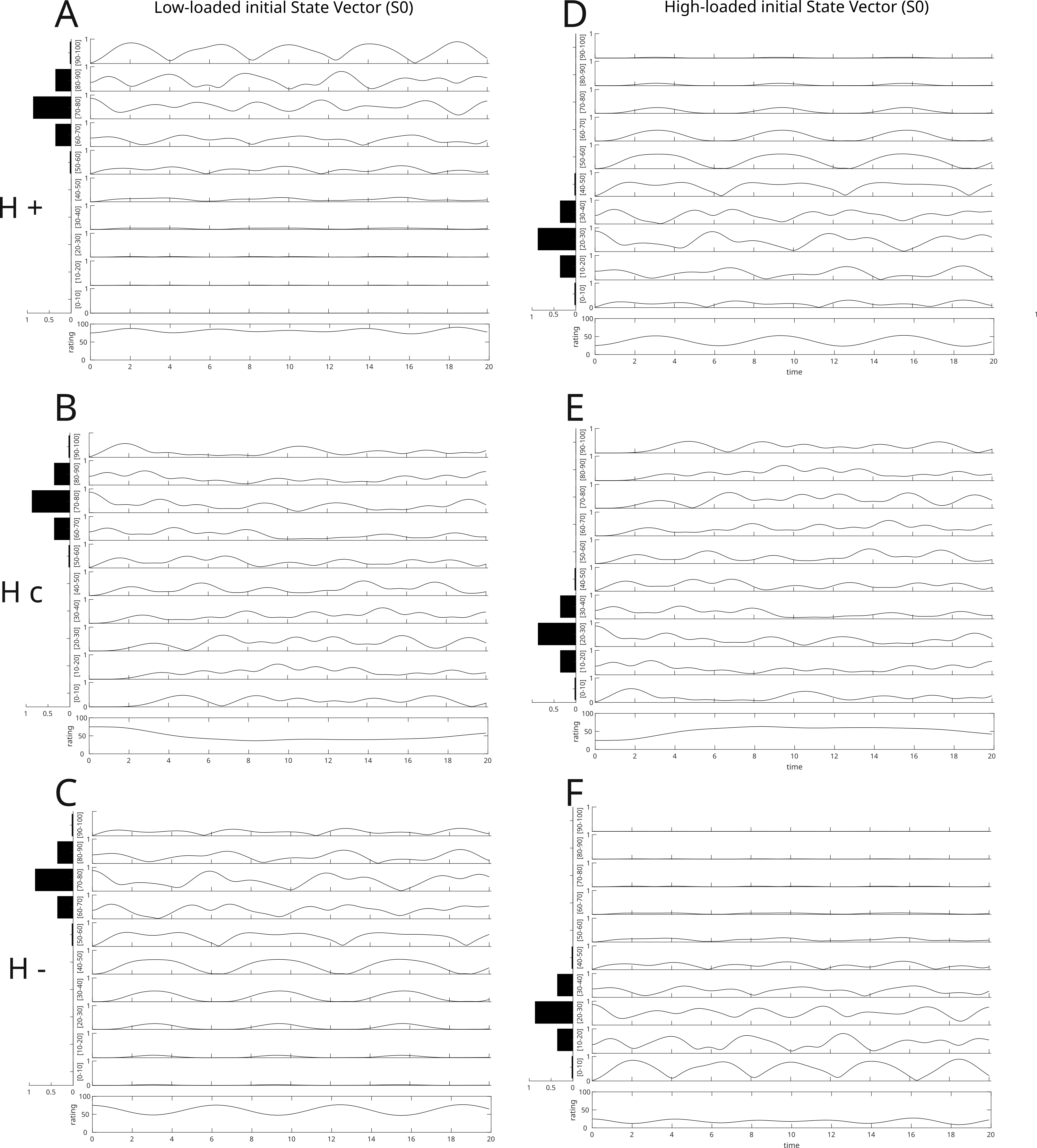}
    \caption{Results of in silico experiments showing the dynamics of the cognitive state amplitudes under different Hamiltonians; Each Panel shows the initial state vector S0 (left vertical bar graph); the temporal evolution of the state propensities for each of the states (central 10 graphs); and finally the modeled reliability rating (bottom single graph). Panels A and B show the evolution using the positive Hamiltonian; Panels C and D with the central Hamiltonian; and Panels E and F with the negative Hamiltonian. The Hamiltonian used impacts which of the states are being explored.}
    \label{fig:insilico}
\end{figure}

In order to explore how the different Hamiltonians distribute the amplitudes in the (initial) state vector over time, we performed several simulations, starting with one state vector with a low initial rating of 2.5 (corresponding to a 25\% reliability rating), and another starting with a state vector of initial rating of 7.5 (corresponding to a 75\% reliability rating). Both are evolved with the positive and negative Hamiltonian (see Figure \ref{fig:insilico}).

These simulations show that a cognitive state vector with propensities for lower reliability ratings states combined with a Hamiltonian with a positive gradient on the diagonal will propel the propensities towards higher reliability ratings. Conversely, a Hamiltonian with a negative gradient will confine exploration towards lower propensities. The Hamiltonian with constant diagonal causes the propensities to evolve throughout all of the 10 reliability ratings. This might indicate that whenever the human agent misses a response, the propensities ``widen out", rather than being driven to lower or higher propensities, which in turn has consequence for the subsequent quantum dynamics. The full effect of using different Hamiltonians to model the reliability according to the subject-specific history can be appreciated with Figure \ref{fig:results_34_and_explanation}, where in Panel B it can clearly be seen that the match-mismatch is distributed differently with each block; and that this has a direct impact on how the cognitive state vector evolves through time. This shows a typical simulation; all other participants yielded similar results. 

\begin{figure}
    \includegraphics[width=0.8\textwidth]{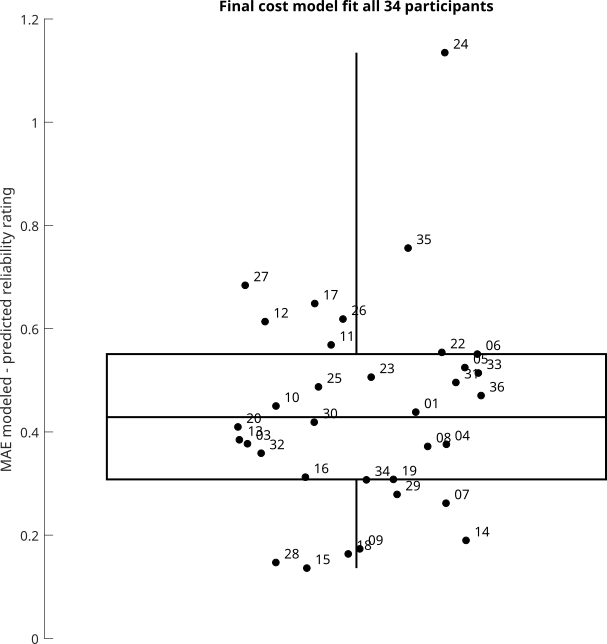}
    \caption{Goodness of fit for all participants, characterized by the mean absolute error. Each participant is marked. The MAE lies between 0.3 and 0.6 (corresponding to 3 and 6 percent on the reliability scape).}
    \label{fig:results_32_and_explanation}
\end{figure}

We fitted the participant-specific QRW model to the data of all 34 participants. While all performed the full experiment, each had a different mix of match and mismatch Hamiltonians and missed trials, and different inter-trial intervals. The quality of fit, characterised by the mean absolute error, is given in Figure \ref{fig:results_32_and_explanation}. Given that the values for the reliability in the model range between 0.5 and 9.5, the QRW model performs reasonably well at predicting the reliability ratings.

\begin{figure}
    \includegraphics[width=0.8\textwidth]{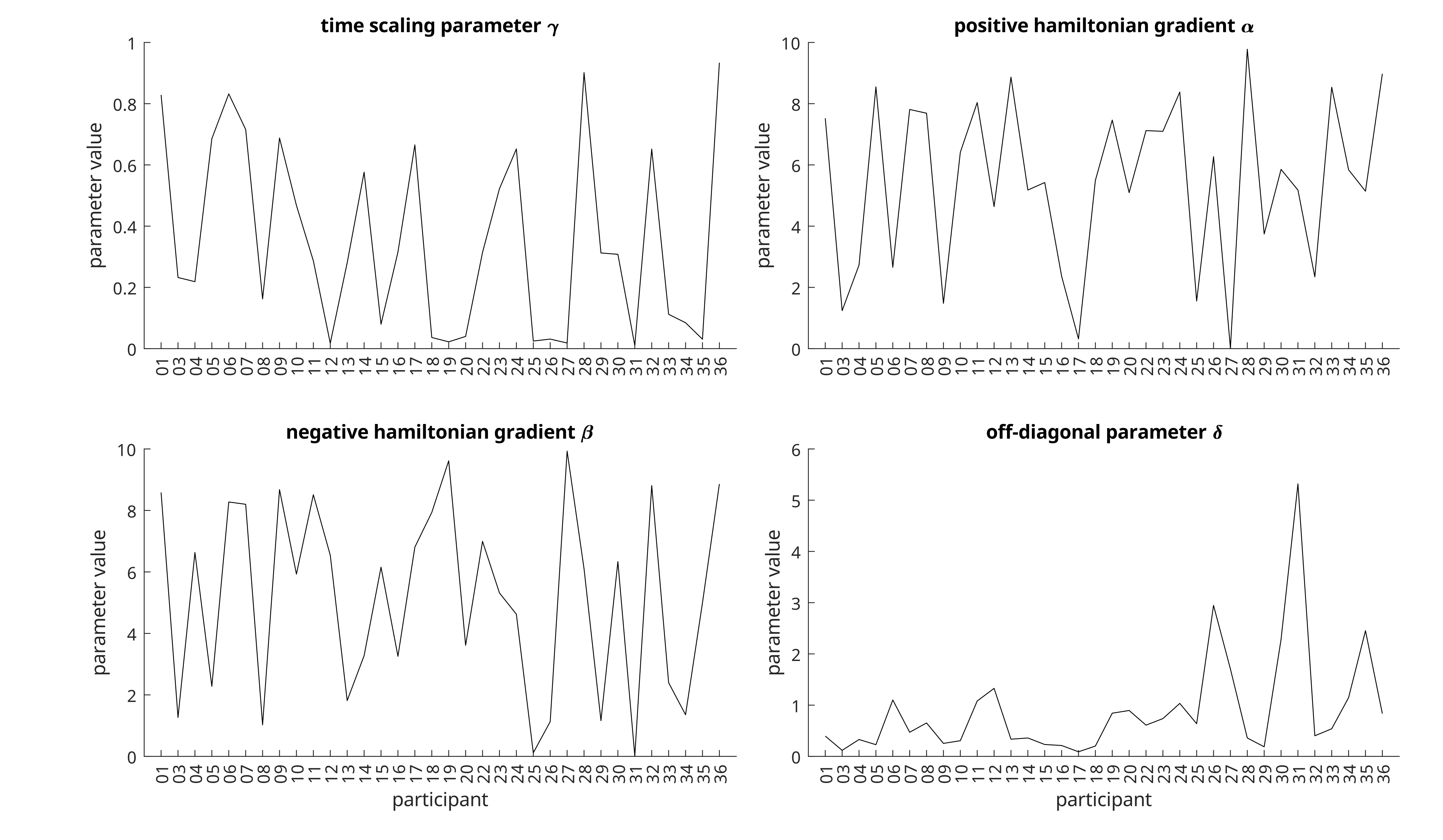}
    \caption{Model parameters for each participant; time scaling $\gamma$; and Hamiltonian positive and negative gradients $\alpha$ and $\beta$, and off-diagonal parameter $\sigma.$}
    \label{fig:drawing_model_parameters}
\end{figure}

We also explore the values of the model parameters themselves in Figure \ref{fig:drawing_model_parameters}. While the time scaling $\gamma$, and the positive and negative gradients $\alpha$ and $\beta$ vary across participants, the $\sigma$ parameter seems to be more similar.

\begin{figure}
    \centering
    \includegraphics[width=0.8\textwidth]{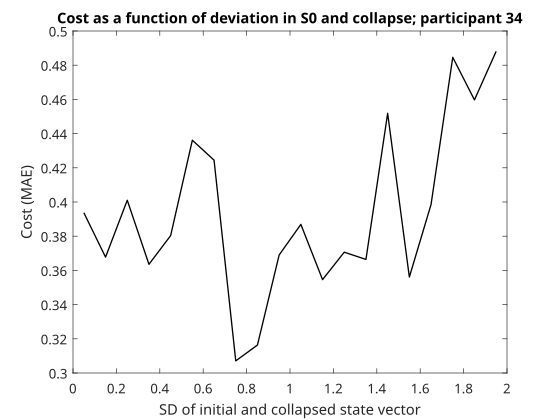}
    \caption{Final mean absolute error (MAE) between model and measurement for different values of the standard deviation used in the initial and collapsed state vectors. In our simulations we used 0.75 (see \ref{fig:insilico}. These simulations are done with participant 34; there is minimal cost between 0.7-0.8; this corresponds to between 7 and 8 \% on the reliability rating.}
    \label{fig:drawing_MAE_as_function_of_cost}
\end{figure}

The standard deviation was not optimized, i.e., the ``spread" used to prime the initial cognitive state vector and to reinitialize it after ``collapse" at the end of each block. The standar deviation was set at a value of 0.75. 
Model simulations were also performed with other values, from 0.05 to 2, in steps of 0.1. See Figure \ref{fig:drawing_MAE_as_function_of_cost} for a comparison of the cost function after optimization is complete. We found that at around 0.8 the cost is minimal.

\section{Discussion}

The adoption of the QRW model for the modeling of the dynamics of reliability ratings over time involves certain modeling choices, particularly in relation to the specification of the Hamiltonian. Each such choice can be reasonably challenged. 
As far as we know, this is the first application of a QRW model where the Hamiltonian is dependent of the interaction between human agent and AI. 
Our aim has been to demonstrate the potential for this approach, particularly in regard to its ability to model the sometimes quite pronounced fluctuations of reliability ratings within a sustained period of human-AI interaction.
With the goal of incorporating a sensitivity in the model to such fluctuations, we have endeavored to make reasonable choices for parameters while keeping the number of modeling parameters low.

The fitting parameters shown in Figure \ref{fig:drawing_model_parameters} appear to vary across participants; however, the off-diagonal parameter $\sigma$ is notably consistent. This unexpected finding warrants further investigation. Our boundary conditions for $\gamma$, $\alpha$, $\beta$, and $\delta$ have been set broadly, ranging from 0.01 to 10. Specifically, for $\alpha$ and $\beta$, we effectively enforce positive and negative gradients along the diagonal of the H+H+ and H-H- Hamiltonians. Whether negative values should also be included or if the upper limit should be higher remains uncertain. These choices were primarily informed by simulations referenced in \cite{busemeyer:bruza:2012}.

In our modeling, we opted for the mean absolute error (MAE) as the cost function instead of the mean squared error (MSE). This decision was driven by the desire to model overall fit without heavily penalizing outliers. Trust development over time is not yet a fully understood mathematical phenomenon (if it were, the MSE might be more appropriate).

Four main modeling assumptions underlie our approach: (1) trust is encoded in a state vector, (2) the state vector collapses when trust or reliability is explicitly queried, (3) evolution depends on the participant's actual timing, and (4) the state vector evolves based on experimental conditions, encoded through different Hamiltonians. A particularly debatable assumption is whether collapse occurs at all, or if it occurs with the standard deviation we used (0.75). Another challenge lies in the Hamiltonians: we use linear gradients in this work, but alternative formulations are possible. For `no response',  we apply a constant value throughout the diagonal of the Hamiltonian, though whether this choice is optimal remains to be seen. Additionally, while we currently model with only "match" and "mismatch" Hamiltonians, and do not differentiate between whether the human judges the image real or fake, it might be argued that mismatches between human-real and AI-fake conditions differ from human-fake and AI-real conditions. This level of differentiation could then justify further refining the model to include four distinct Hamiltonians.

Our results indicate that the mix of trials (and consequently, the mix of positive, negative, and neutral Hamiltonians) significantly influences the modeled dynamics of reliability. Given the close alignment between modeled and measured values, we posit that events and choices in human-AI teaming play a critical role in shaping judgments of reliability and that this process is somewhat adequately captured by our Hamiltonians.


The role of Hamiltonians in representing the evolution of cognitive states brain states may provide connections to the neuroscience literature examining "oddball" stimuli (less frequent events) compared to "normal" stimuli, which often elicit distinct encephalography (EEG) responses \cite{polich:2012,luck2014introduction}. 
From a neuroscience perspective, this suggests the existence of different brain states, as neuroimaging shows distinct signals at various brain locations during these trials. In our model, one brain state corresponds to processing a "match" between human and AI, while another processes a "mismatch." However, from a modeling perspective, these brain states act as recipes for evolving an internal, indeterminate evaluation of trust, described by a state vector. Future work could explore integrating short-term brain states (measured via neuroimaging) with the model over time to account for the longer-term processing of trust.

Modeling internal brain processes helps us understand the mechanisms behind cognition, going beyond measurements to predict and explain behavior. A QRW model is a strong candidate because it captures the probabilistic and dynamic nature of decision-making, aligning with how trust or uncertainty evolves over time. However, other quantum dynamical models, such as spin systems, might also be worth exploring. Spin systems are curiously analogous to the experiment in the sense that the judgments of real/fake can be viewed as spin measurements up/down in relation to two interacting spin systems namely the human agent and the AI.

\section{Future Directions}

The QRW presented above depends on whether the interaction with AI aligns with the human agent or not. Recall that when there is alignment, the Hamiltonian has a diagonal with positively increasing values. Conversely, if there is mis-alignment, the diagonal is primed with negatively increasing values. A future research direction would be to use real-time EEG analysis to determine which Hamiltonian to use. In a previous study \cite{roeder2023quantum}, we showed that there is a statistically significant different event-related potentials (ERP) evident in the EEG when the AI disagrees with human judgment, compared to when it does. If the difference between ERPs can be determined in real-time, then the appropriate Hamiltonian could be used to drive the dynamics of the QRW model, which in turn can be used to compute the human agent's predictions of AI's reliability in future interactions. These predictions will be especially important in situations where the reliability is predicted to decrease, because once human agents lose trust in AI, it is challenging to repair it \cite{deVisser:2018:trust-repair}. With light-weight, high fidelity EEG head-sets becoming available, one can envisage a system where human agents wear these head-sets when interacting with AI and when the human judgment of reliability is predicted to significantly decrease. In such a case, a suitable intervention could be triggered to ameliorate the AI's functioning in order to prevent an associated loss of trust.

\subsection*{Acknowledgments}
This research was supported by the Air Force Office of Scientific Research under award number: FA9550-23-1-0258.
\bibliographystyle{RS}
\bibliography{references}
\end{document}